\documentclass[a4paper,11pt]{article}
\usepackage{jinstpub} 
\usepackage{lineno}

\title{\boldmath Updated parameters of the LArQL model}

\author[a,1]{L. Paulucci,\note{Corresponding author.}}
\author[b]{F. Cavanna,}
\author[c]{V. Vale}
\author[c]{and F. Marinho}
\affiliation[a]{Universidade Federal do ABC,\\Santo André, SP, 09210-170, Brazil}
\affiliation[b]{Fermi National Accelerator Laboratory,\\ Batavia, IL 60510, U.S.A.}
\affiliation[c]{Instituto Tecnológico de Aeronáutica,\\São José dos Campos, SP, 12228-900, Brazil}

\emailAdd{laura.paulucci@ufabc.edu.br}

\abstract{The need for a microscopic description of scintillation light generation in liquid argon becomes increasingly desirable with the upcoming operation of large scale LArTPCs in the next decade. While a detailed mathematical account of the process is still to be achieved, a phenomenological model for simultaneously treating ionization and scintillation, LArQL, has been successfully employed to describe the range of electric fields from 0 to 0.75 kV/cm and dE/dx from 2 to 40 MeV/cm providing the anti-correlation between the free ionization charge and scintillation light. A reanalysis of the original model parameter values has been performed within a global fit procedure and is presented.}


\begin{document}
\maketitle
\flushbottom

\section{Introduction}
\label{sec:intro}

There are many open questions in neutrino physics such as what is the neutrino mass ordering, CP violation in the leptonic sector, and the possible existence of a fourth neutrino flavor. A number of projects aim to investigate in details the properties of neutrinos using different detection techniques. One of those is the use of time projection chambers in large liquid argon volumes (LArTPCs) given their high precision imaging capabilities.

In a LArTPC an electric field is applied in order to collect the ionization electrons that are created along a charged particle trajectory. Those pass through a set of induction planes before being collected by the anode, allowing for a 3D reconstruction of the particle’s trajectory with precision on the millimeters range.  

In addition to collecting electrons from ionization, argon scintillation light is also collected and used for fast timing of events. It can also be used for calorimetric reconstruction if the light collection is done in an efficient manner. The combination of TPC collected charge and scintillation light has been used to improve the energy resolution in EXO~\cite{exo}, a xenon TPC experiment.

Energy reconstruction is critical for neutrino oscillation analysis. The energy deposited in liquid argon will be split into ionizing electrons and scintillation photons. Moreover, there is an anti-correlation between the two quantities with a clear dependency on the applied electric field~\cite{kubota, doke1981, doke1988, Doke2002}. Therefore, a LArTPC does not measure all the energy deposited by charged particles due to the recombination factor $R$. 


\section{LArQL model}

LArQL~\cite{Larql} is an unitary model for Ionization Charge ($Q$) and Scintillation Light ($L$) in LAr, and considers both quantities as functions of the particle's deposited energy density (dE/dx) and local electric field ($\mathscr{E}$). Its validity is evaluated in the ranges: 0 kV/cm < $\mathscr{E}$ < 0.75 kV/cm and 2 MeV/cm < $dE/dx$ < 40 MeV/cm, spanning over the operation range for LArTPC neutrino experiments.
The model is built with a charge-light master equation:
 \begin{equation}\label{ME}
     Q+L=~N_i~+~N_{ex},
 \end{equation}
where $Q$ and $L$ are parametrized as functions with a minimal number of parameters to be fitted to data and the quantities:
\begin{enumerate}
    \item $N_{i} = 1/W_{ion}$ :  ionizations per energy unit;
    \item $N_{ex}/N_{i}$:  excitations/ionizations;
    \item $\chi_0$(dE/dx): fraction of escaping electrons.     
\end{enumerate}

The LArQL model is developed from the Birks' charge recombination model with the addition of escaping electrons, electrons which escape recombination with Ar ions even when the external electric field applied is zero. The fraction of escaping electrons, $\chi_{0}$, can be obtained from experimental data showing the reduction of the light yield at zero field ($\eta_0$) for lower $dE/dx$ values. If we consider the fraction of missing photons to be written as ($1 - \eta_0$), then

\begin{equation}\label{eta0}
    \chi_{0} = (1+N_{ex}/N_i) \cdot (1-\eta_0).
\end{equation}

In the Birks’ charge parametrization, the ionization charge goes to zero when there is no electric field actively drifting electrons away from their parent ion, which is in disagreement with data indicating the existence of escape electrons~\cite{Doke2002}. At low electric field and any $dE/dx$, escaping electrons should be considered.  Therefore, a correction term to the Birks’ charge model is introduced in LArQL as:
\begin{equation}\label{qy}
Q_{LArQL} = \frac{A_B/W_{ion}}{1 + \frac{k_B}{\rho_{LAr}}\cdot \frac{1}{\mathscr{E}}\cdot \frac{dE}{dx}} + \chi_0\left({dE/dx}\right) \cdot f_{corr}(\mathscr{E},dE/dx)\cdot Q_{\infty},
\end{equation}
where the first term on the right is the Birks' formula with its $A_B$, $k_B$ original parameters and LAr density ($\rho_{LAr}$)~\cite{icarus}. The second term accounts for the escape electrons and is given by the product of the function $\chi_0$, which describes the fraction of escaping electrons at $\mathscr{E} = 0$, a factor $f_{corr}$ introducing adequate $\mathscr{E}$ dependence, and the maximum charge yield, $Q_{\infty}$, expected at high $\mathscr{E}$ regime. Both $\chi_0$ and $f_{corr}$ are phenomenological, fitted to data, and their functional forms are
\begin{equation}
    f_{corr} = e^{-\mathscr{E}/(\alpha \ln dE/dx + \beta) }, \,\,\,\,\,\, \chi_0=A / [B + e^{(C + D\, dE/dx)}],
\end{equation}
Therefore in this model additional six free parameters are introduced ($\alpha$, $\beta$, $A$, $B$, $C$, and $D$). The pair ($\alpha$, $\beta$) regards the $\mathscr{E}$ correction and ($A$, $B$, $C$, $D$) allow for the evaluation of amount of escaping electrons at null electric field.
Parameters of the model have been adjusted to fit available data, both for charge and light yield.


\section{Model Regression}

We have performed a global statistical fit, weighted by available data sets. Ionization charge data was obtained from ref.~\cite{icarus} while scintillation light data was taken from refs.~\cite{Doke2002, Aris}, the same set as used in our previous work~\cite{Larql}. After determining the best fit parameters, results were compared with data from ref.~\cite{Microboone}.

The code was developed in C++, using the ROOT framework, to explore variations of LArQL parameters, including those from Birks model ($A_B$ and $k_B$), limited to their known experimental uncertainties. With this procedure, a new fit was achieved. This global fit was performed minimizing the sum of the weighted residual squares for charge and light experimental datasets simultaneously. 

The parameters value ranges used for the minimization procedure were first narrowed down through broader multidimensional scans studies. It was verified that the best region to perform searches should be around the values originally adopted in the LArQL model~\cite{Larql} and also allowing $A_B$ and $k_B$ to vary within their $\rm 3\sigma$ confidence level. On the other hand, when adopting large variations for the parameters with respect to their original values it was found that in certain conditions only some of the data sets could be described with similar quality while for others, estimates degraded.

The weights were evaluated using the $RMS_{best}^i$ obtained from an intermediate fit of the model to each {\it i-th} dataset individually, providing best estimates for these RMSs given LArQL satisfactory data description. 

\begin{figure}[htbp]
\centering
\includegraphics[width=.45\textwidth]{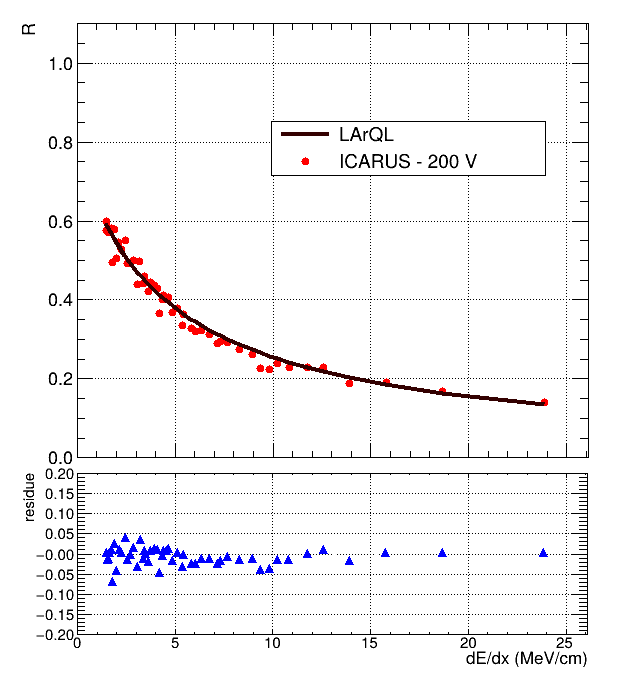}
\includegraphics[width=.435\textwidth]{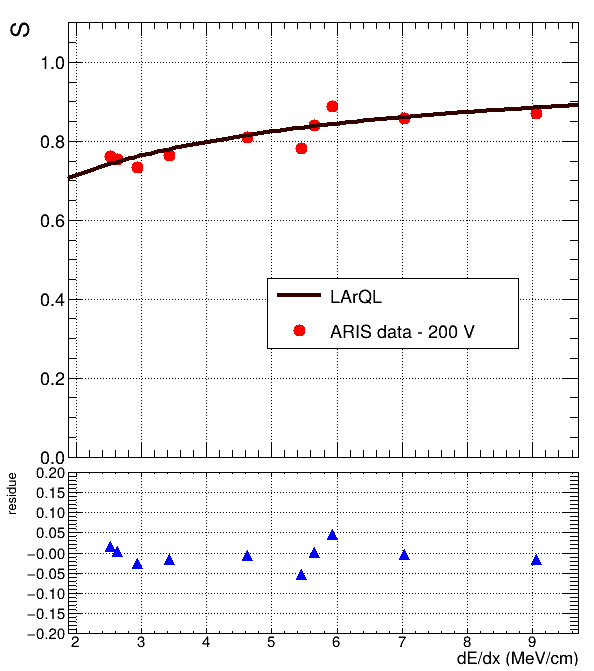}
\caption{Comparison between data (dots) and LArQL (curve) on the top frame and their respective residuals at the bottom frame as a function of $dE/dx$ at $\mathscr{E}\rm = 200 V/cm$. On the left side panel, data for the recombination factor $R$ obtained by the ICARUS collaboration~\cite{icarus} is shown while on the right panel we present data for $S$, the ratio between scintillation at the given electrical field and zero field, from the ARIS collaboration~\cite{Aris}.\label{fig:resid}}
\end{figure}

Assuming the residuals for all data groups are normal distributed, minimization of the weighted sum of squared residuals allows for a satisfactory estimate of the parameters:

\begin{equation}
WSSR = \sum_i \frac {SSR_i}{n_i (RMS^i_{best})^2},
\end{equation}
where $SSR_i$ is the sum of squared residuals for the {\it i-th} data set and $n_i$ is the number of measurements of that set. Figure \ref{fig:resid} shows examples of comparison plots between data and LArQL and their respective residuals at the bottom frame both as functions of $dE/dx$. The recombination factor $R$ data from \cite{icarus} is seen on the left side panel while $S1/S1_0$ from \cite{Aris} is on the right panel for both experiments at $\mathscr{E}\rm = 200 V/cm$. The same procedure was performed for data from the same experiments at other configurations of electric fields and similar descriptive quality was achieved with LArQL. All $SSR_i$s for each dataset were individually reduced with respect to the original LArQL parameters set\cite{Larql}.

The parameters optimization was achieved via a simple search algorithm using independent multidimensional random samplings of the parameter values. Parameters were sampled uniformly and independently across their allowed ranges. This fit allowed to achieve better residuals for all data sets in comparison with the original version of the model. All $SSR^i_{fit}$ values obtained through the global fit are relatively smaller than the obtained with the original LArQL set of parameters ($|1-SSR^i_{fit}/SSR^i_{original}|$ <10\%), indicating that the global fit provides an overall improved general description of all data sets.

Table~\ref{tab:Params} summarizes the most probable parameter values obtained from the LArQL model optimization. On the original version of the model it was assumed the values for $A_B$ and $k_B$ fixed as proposed in \cite{icarus} while the parameters related to the term accounting to escaping electrons and its $\mathscr{E}$ dependence were adapted accordingly to provide satisfactory estimates. The $A_B$ and $k_B$ uncertainties values indicated are the same as from \cite{icarus} while the remaining parameters had only central value estimates. On the procedure presented in this work it was possible to stablish a fit to optimize the model considering all data sets used simultaneously and with same weight. Uncertainties were also evaluated for all parameters. Notice, the central values and uncertainties for $A_B$ and $k_B$ obtained from the global fit are not to be interpreted as new or better estimates for the Birks formula alone and can only be reasonably employed in the full LArQL model. 


\begin{table}[h]
\centering
\caption{Comparison of the original parameter values presented in ref.~\cite{Larql} for the LArQL model and the new parameters after the global fit procedure. The uncertainties of the parameters were obtained together with the appropriately considered statistic correlations.\label{tab:Params}}
\smallskip
\begin{tabular}{|l|c|c|c|c|}
\hline
                      & {$A_B$}    & {\begin{tabular}[c]{@{}c@{}}$k_B$ {\small(gV/MeV cm$^3$})\end{tabular}} & {\begin{tabular}[c]{@{}c@{}}$\alpha$ {\small(cm/kV)}\end{tabular}} & {\begin{tabular}[c]{@{}c@{}}$\beta$ {\small (cm/kV)}\end{tabular}} \\ \hline
{Original} & 0.800(3)             & 48.6(6)                                                                & 0.0372                                                              & 0.0124                                                             \\ \hline
{Global fit}      & 0.808(2)          & 49.7(4)                                                             & 0.0387(8)                                                           & 0.0128(6)                                                          \\ \hline \hline
                      & {\it A}        & {\it B}                                                          & {\it C}                                                          & {\begin{tabular}[c]{@{}c@{}}{\it D} {\small(cm/MeV)}\end{tabular}}      \\ \hline
{Original} & $3.38.10^{-3}$    & -6.6                                                                & 1.88                                                                & $1.29.10^{-4}$                                                     \\ \hline
{Global fit}      & $3.61(5).10^{-3}$ & -5.7(1)                                                             & 1.74(2)                                                             & $2.01(3).10^{-4}$                                                  \\ \hline
\end{tabular}
\end{table}

\begin{figure}[htbp]
\centering
\includegraphics[width=.7\textwidth]{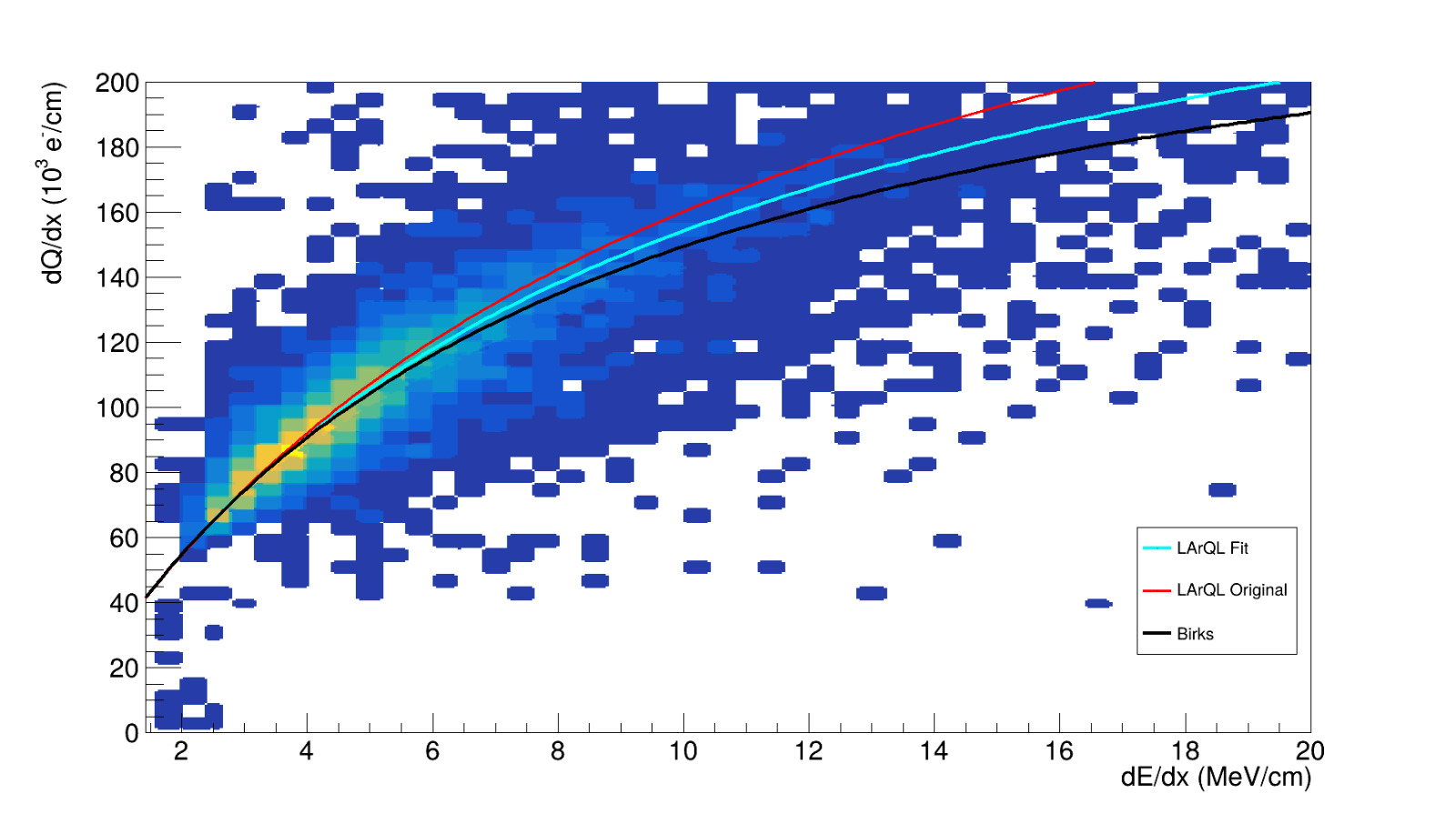}
\includegraphics[width=.7\textwidth]{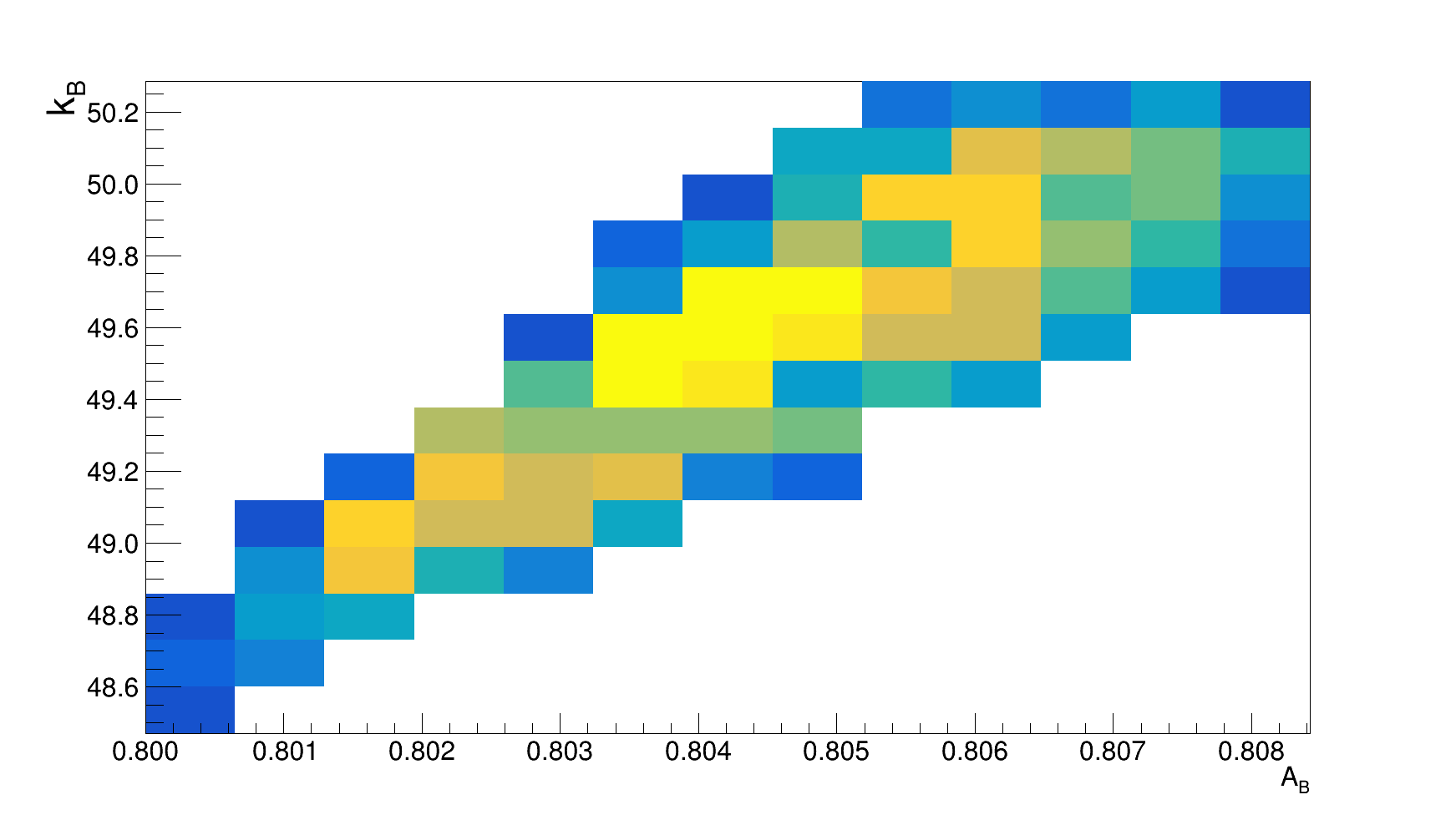}
\caption{(Left) Microboone dQ/dx data~\cite{Microboone} with Birks in black and two LArQL lines: in red with the original parameters~\cite{Larql} and in cyan with the parameters obtained in this work. The improvement in agreement of the latter is visible. (Right) Correlation scatter plot between $A_B$ and $k_B$ parameters on LArQL. \label{fig:Comp1}}\end{figure}

The left panel of fig.~\ref{fig:Comp1} shows measurements of $dQ/dx$ as function of $dE/dx$ from the Microboone experiment \cite{Microboone} and correspondent curves predicted by the Birks model (black) and LArQL with the original set of parameters (red) and with the latest parameters from the global fit (cyan). Both instances of LArQL (red and cyan) provide satisfactory estimates while the Birks model slightly under estimates $dQ/dx$ at higher $dE/dx$ values.

The correlations between the fitted parameters were also investigated through typical 2D scatter plots and a number of features were noticed. For instance, $A_B$ and $k_B$ are strongly correlated (see figure \ref{fig:Comp1}) while both these parameters do not exhibit any clear constraints to the (A, B, C and D) set. The independent variation of $A_B$($k_B$) with fixing $k_B$($A_B$) value impacts the electric field dependence of the model degrading its data description capability. On the other hand, $\alpha$ is also correlated to both $A_B$ and $k_B$, perhaps, because it dictates the $\mathscr{E}$ dependence of the escaping electrons term. Otherwise, it is also fairly unrelated to any other LArQL parameter. The $\beta$ parameter seems to be uncorrelated to all the other parameters within the its range of considered values. The (A, B, C, D) set of parameters are somewhat constraint between themselves but are independent of the other model parameters.

\section{Conclusions and Future Prospects}

In a LArTPC the deposited energy can originate either ionization electrons or scintillation photons in an anti-correlated manner. In order to describe the combined production of ionization charge and scintillation light a phenomenological model, LArQL, has been developed. Here we presented a redefinition of the model’s parameters values using a global fit procedure.
This model has been made available in the LArSoft framework~\cite{larsoft}, commonly used by LArTPC experiments.

We aim at further improve the charge and light generation modeling with LArQL by including more available datasets on the analysis and exploring other regimes of particle energy deposition. For that, a more robust regression model should be developed in order to provide a more efficient optimization and allow for accounting of any specific details of the different experiments data sets considered.




\bibliographystyle{JHEP}
\bibliography{biblio.bib}

\providecommand{\href}[2]{#2}\begingroup\raggedright\begin{thebibliography}{10}

\bibitem{exo}
E.~Conti et~al., \emph{{Correlated fluctuations between luminescence and ionization in liquid xenon}}, \href{https://doi.org/10.1103/PhysRevB.68.054201}{\emph{Phys. Rev. B} {\bfseries 68} (2003) 054201} [\href{https://arxiv.org/abs/hep-ex/0303008}{{\ttfamily hep-ex/0303008}}].

\bibitem{kubota}
S.~Kubota et~al., \emph{Recombination luminescence in liquid argon and in liquid xenon}, {\emph{Phys. Rev. B} {\bfseries 17} (1978) 2762}.

\bibitem{doke1981}
T.~Doke, \emph{Fundamental properties of liquid argon, krypton and xenon as radiation detector media}, {\emph{Portugal Phys.} {\bfseries 12} (1981) 9}.

\bibitem{doke1988}
T.~Doke et~al., \emph{Let dependence of scintillation yields in liquid argon}, {\emph{Nucl. Instrum. Meth. A} {\bfseries 269} (1988) 291}.

\bibitem{Doke2002}
T.~Doke, A.~Hitachi, J.~Kikuchi, K.~Masuda, H.~Okada and E.~Shibamura, \emph{{Absolute scintillation yields in liquid argon and xenon for various particles}}, \href{https://doi.org/DOI}{\emph{Jpn. J. Appl. Phys.} {\bfseries 41} (2002) 1538}.

\bibitem{Larql}
F.~Marinho, L.~Paulucci, D.~Totani and F.~Cavanna, \emph{{LArQL: a phenomenological model for treating light and charge generation in liquid argon}}, \href{https://doi.org/DOI}{\emph{JINST} {\bfseries 17} (2022) C07009}.

\bibitem{icarus}
S.~Amoruso et~al., \emph{{Study of electron recombination in liquid argon with the ICARUS TPC}}, \href{https://doi.org/DOI}{\emph{Nucl. Instrum. \& Methods A} {\bfseries 523} (2004) 275}.

\bibitem{Aris}
P.~Agnes et~al., \emph{{Measurement of the liquid argon energy response to nuclear and electronic recoils}}, \href{https://doi.org/10.48550/arXiv.1801.06653}{\emph{Phys. Rev. D} {\bfseries 97} (2018) 112005} [\href{https://arxiv.org/abs/1801.06653}{{\ttfamily 1801.06653}}].

\bibitem{Microboone}
C.~Adams et~al., \emph{{Calibration of the charge and energy loss per unit length of the MicroBooNE liquid argon time projection chamber using muons and protons}}, \href{https://doi.org/10.48550/arXiv.1907.11736}{\emph{JINST} {\bfseries 15} (2020) P03022} [\href{https://arxiv.org/abs/1907.11736}{{\ttfamily 1907.11736}}].

\bibitem{larsoft}
E.D.~Church, ``Larsoft: A software package for liquid argon time projection drift chambers.'' arXiv:1311.6774 [physics.ins-det], \url{https://larsoft.org/}.

\end{thebibliography}\endgroup




\end{document}